# Unlocking new capabilities in the analysis of GC×GC-TOFMS data with shift-invariant multi-linearity

## Authors

Paul-Albert Schneide[1,2], Michael Sochoran Armstrong[3], Neal Gallagher[4], Rasmus Bro[1]

[1] Department of Food Science, University of Copenhagen, Frederiksberg, Denmark

[2] Department of Analytical Science, BASF SE, Ludwigshafen am Rhein, Rhineland-Palatinate, Germany

[3] Departamento de Teoría de la Señal, Telemática y Comunicaciones, Universidad de Granada, Spain

[4] Eigenvector Research, Inc., Manson, Washington, USA

## Abstract

This paper introduces a novel deconvolution algorithm, shift-invariant multi-linearity (SIML), which significantly enhances the analysis of data from a comprehensive two-dimensional gas chromatograph coupled to a mass spectrometric detector (GC×GC-TOFMS). Designed to address the challenges posed by retention time shifts and high noise levels, SIML incorporates wavelet-based smoothing and Fourier-Transform based shift-correction within the multivariate curve resolution-alternating least squares (MCR-ALS) framework. We benchmarked the SIML algorithm against traditional methods such as MCR-ALS and Parallel Factor Analysis 2 with flexible coupling (PARAFAC2×N) using both simulated and real GC×GC-TOFMS datasets. Our results demonstrate that SIML provides unique solutions with significantly improved robustness, particularly in low signal-to-noise ratio scenarios, where it maintains high accuracy in estimating mass spectra and concentrations. The enhanced reliability of quantitative analyses afforded by SIML underscores its potential for broad application in complex matrix analyses across environmental science, food chemistry, and biological research.

# Abbreviations

| | |
|---|---|
| $^1$D | First retention dimension in two-dimensional chromatography |
| $^2$D | Second retention dimension in two-dimensional chromatography |
| ALS | Alternating Least Squares |
| BPC | Base peak chromatogram |
| CLS | Classical least squares |
| FFT | Fast Fourier Transform |
| FT | Fourier Transform |
| GC×GC-TOFMS | Two dimensional gas chromatography coupled to a mass spectrometric (time-of-flight) detector |
| MCR | Multivariate curve resolution |
| NMF | Non-negative matrix factorization |
| PARAFAC | Parallel factor analysis |
| PARAFAC2 | Parallel factor analysis 2 |
| PARAFAC2xN | Parallel factor analysis 2 with flexible coupling constraint allowing for shift in more than one mode |
| SIML | Shift-invariant multi-linearity algorithm |
| SIML-DN | Shift-invariant multi-linearity algorithm with denoising |
| SNR | Signal-to-noise ratio |
| SSE | Sum of squared errors |
| SST | Total sum of squares |
| TIC | Total ion chromatogram |
| SVD | Singular value decomposition |
| TMS | Tri-methyl-silyl (protective group) |

## Notation

$\underline{X}^{(p)}$     $p$th order tensor

$X$     ($I \times J$) matrix, equivalent to $\underline{X}^{(2)}$

$x$     ($I \times 1$) vector, equivalent to $\underline{X}^{(1)}$

$x_{ijk}$     Scalar, equivalent to $\underline{X}^{(0)}$

$T$     ($1 \times I$) transpose of $x$

$\odot$     Kathri-Rao product (column-wise Kronecker product)

$\circ$     Element-wise matrix multiplication (Hadamard product)

$||\cdot||_F^2$     Frobenius Norm

$\hat{f}(k)$     Fourier transform of $f(n)$

$\hat{f}^*(k)$     Complex conjugate of $\hat{f}(k)$

$|\hat{f}(k)|$     Amplitude spectrum of $f(n)$

## 1. Introduction

Comprehensive two-dimensional gas chromatography time-of-flight mass spectrometry (GC×GC-TOFMS) is a powerful analytical technique that allows for comprehensive separation and characterization of very complex samples such as pyrolysis oils, environmental samples, biological, and food samples [1–5]. However, handling the data is challenging and time consuming. The challenge is particularly evident in exploratory, untargeted analysis where the aim is to obtain an exhaustive chemical fingerprint of the sample composition.[6–8] A recently published benchmark of eight different commercial and open-source software packages revealed larger differences in the data processing capabilities of the individual software packages, most notably with respect to the identification of features in untargeted analysis.[9] In addition to the missing standardization in the data processing workflows of available software tools, their functionality is in some cases not sufficient, which fuels the development of novel algorithms for more sophisticated chemical information extraction (deconvolution), pattern recognition, or down stream statistical analysis.[7,8,10–13] Chemometric methods such as Multivariate Curve Resolution [MCR[14–16], also known as non-negative matrix factorization (NMF)[17–19] in the data science community], Parallel Factor Analysis (PARAFAC) and extended versions of Parallel Factor Analysis 2 (PARAFAC2×N) have been described as useful methods for deconvolution in targeted and untargeted GC×GC-TOFMS data analysis.[13,20,21] Nevertheless, there are limitations associated with each of the currently known deconvolution algorithms in their application to GC×GC-TOFMS data analysis. For example, the structure of the MCR model accounts for retention time shifts occurring in the first and second retention dimension; however, MCR does not generally provide unique solutions.[22,23] The rotational ambiguity of MCR solutions can cause large variabilities in the estimated qualitative and quantitative information.[24] On the other hand, PARAFAC and PARAFAC2 provide unique solutions but have higher requirements on the data structure. In the case of PARAFAC, the GC×GC-TOFMS data needs to be perfectly aligned. Unfortunately, retention time shifts violate the PARAFAC model assumptions.[25] The PARAFAC2×N method is more flexible and can model data that deviates from a multi-linear structure caused by retention time shifting but comes with a high algorithmic complexity.[13,26]

In this paper a new deconvolution algorithm called shift-invariant multi-linearity (SIML) is presented. The algorithm is inspired by the multi-linearity constraint proposed by Tauler et al. and the shift-invariant tri-linearity constraint proposed by Schneide et al.[27,28] The proposed method integrates a wavelet-based smoothing and the shift-invariance properties of the Fourier-Transform into the MCR-ALS routine to effectively correct intra- and inter-sample shifts yielding unique solutions. The algorithm is benchmarked against MCR-ALS and PARAFAC2×N on challenging simulated and real multi-sample GC×GC-TOFMS data sets.

## 2. Background

### 2.1 Data structure

The data structure of a single GC-MS measurement can be described as a matrix $X$ with dimensions ($I\times J$), with $I$ denoting the number of scans in the first retention dimension ($^1$D) and $J$ denoting the mass scans (mz). In extension to that, a GC×GC-TOFMS measurement naturally has the form of a third order tensor $\underline{X}^{(3)}$ with dimensions ($I\times K\times J$) in which $K$ describes the scans in the second retention dimension ($^2$D). One practical way of visualizing $\underline{X}^{(3)}$ is thinking of it as $K$ slices $\underline{X}^{(2)}$ with dimensions ($I\times J$) or as $I$ slices $\underline{X}^{(2)}$ of dimension ($K\times J$). Thus, a GC×GC-TOFMS measurement can also be expressed in the form of $K$ concatenated slices $\underline{X}^{(2)}$ with dimensions ($I\times J$), giving an augmented matrix $X$ with dimensions ($IK\times J$). This idea also extends to the situation of having a set of several GC×GC-TOFMS measurements, which can be arranged to form a fourth order tensor $\underline{X}^{(4)}$ with dimensions ($I\times K\times L\times J$) or an augmented matrix $X$ with dimensions ($IKL\times J$), where $L$ is the number of samples.

To aid visualization, examples of the data structures for the case of a single GC×GC-TOFMS measurement and multiple GC×GC-TOFMS measurements are given in Figure 1.

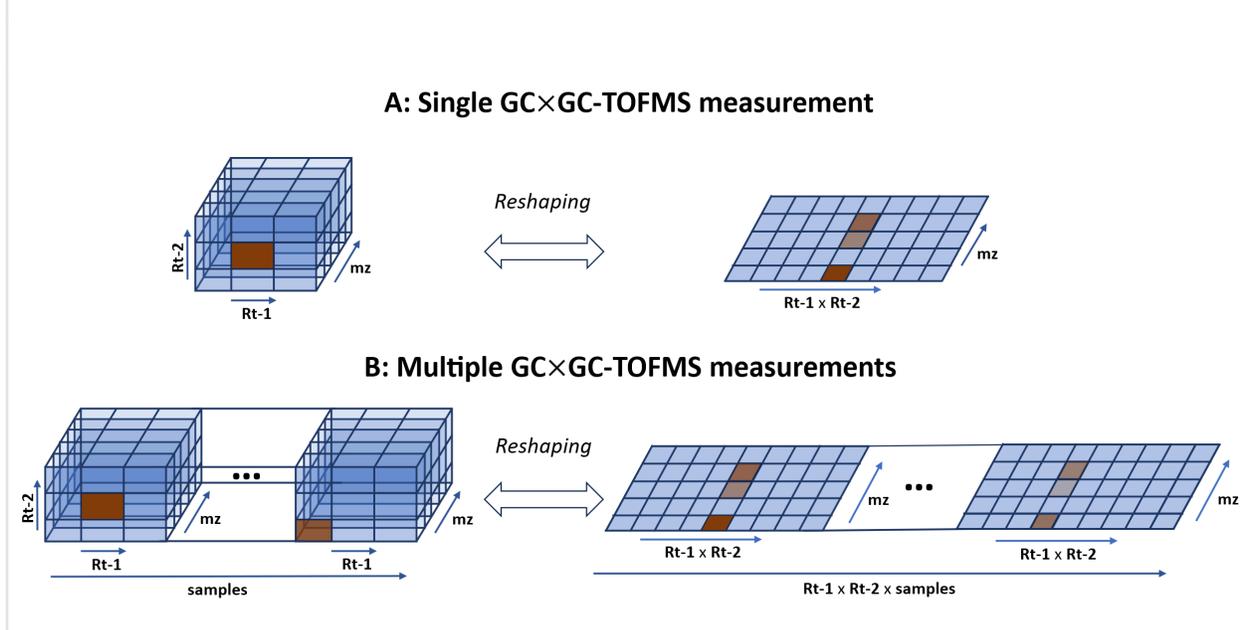

*Figure 1: Visualization of the data structure of **A:** a single GC×GC-TOFMS measurement organized as higher order tensor or as augmented matrix and **B:** a set of multiple GC×GC-TOFMS measurements organized as higher order tensor or as augmented matrix.*

## 2.2 Algorithms for modeling GC×GC-TOFMS data

Different chemometric approaches have been described for extracting quantitative (concentrations) and qualitative (mass spectra) information from GC×GC-TOFMS data. Specifically, MCR-ALS, PARAFAC and PARAFAC2×N were described in literature and will be explained in the following.

Multivariate Curve Resolution is a bilinear factorization method which decomposes a matrix $X$ into two positive matrices $C$ and $S$. In the context of GC×GC-TOFMS data analysis, $X$ is the set of unfolded GC×GC-TOFMS measurements with dimensions $(IKL \times J)$, $C$ is a factor matrix with dimensions $(IKL \times R)$ containing the concatenated elution profiles, and $S$ is a $(J \times R)$ sized factor matrix holding the analyte mass spectra. The MCR model can be formulated according to Equation 1:

Equation 1: $$X = CS^T + E$$

Different algorithms have been proposed to calculate $C$ and $S$ but this paper will focus on the most prominently used Alternating Least Squares (ALS) algorithm. The loss function for the ALS algorithm with non-negativity constraints[29] can be formulated according to Equation 2:

Equation 2: $$\left\| X - CS^T \right\|_F^2 \text{ s.t.}$$
$$C_{nr} \geq 0 \ \forall \ n \in \{1, ..., IKL\}, \ r \in \{1, ..., R\}$$
$$S_{mr} \geq 0 \ \forall \ m \in \{1, ..., J\}, \ r \in \{1, ..., R\}$$

A major advantage of MCR is that it can deconvolve overlapped signals and effectively model chromatographic artifacts such as retention time shifts and changes in peak shape. But a major disadvantage of MCR is that it suffers from a rotational ambiguity which means that a range of solutions for $C$ and $S$ exist that all minimize $L(C, S)$. Mathematically, this can be shown by Equation 3 in which $C$ and $S$ are solutions obtained by fitting the MCR model with one set of initial values and $C_A$ and $S_A$ are rotated solutions (fulfilling the applied constraints) that would provide the same fit. The matrix $T$ is a rotation matrix, such that the matrix product $TT^{-1}$ becomes identity.

Equation 3:
$$C_A S_A^T = (CT)(T^{-1}S^T)$$

Several scientific works investigated methods for estimating the range of feasible solutions theoretically or practically to derive estimates on how well-defined a given solution is.[22,24,30,31] It is intuitively clear that the range of feasible solutions can be reduced by constraining $L(C, S)$ and several constraints have been proposed to reduce the range of feasible solutions utilizing a priori knowledge about the measurement principle and the data characteristics. Specifically, the multi-linearity constraint[27] can provide unique solutions for data structures $\underline{X}^{(p \geq 3)}$ (e.g., GC×GC-TOFMS data)[27], and can be seen as a particular implementation of the PARAFAC/CANDECOMP model.[32,33] However, multi-linearity is a strong constraint that requires elution profiles of a given analyte to remain constant in shape and position across modulations, and sample-to-sample.[25] These conditions for multi-linearity are not always satisfied for GC×GC-TOFMS data, most importantly because of retention time shifts. Although the application of PARAFAC for the decomposition of single GC×GC-TOFMS measurements has been described in literature[21], this approach will only be valid in the absence of shift between the different modulations. if analytes elute over a short time window in $^1$D or if the temperature in $^2$D is held constant.[7] If instead the temperature in $^2$D is ramped up alongside increasing number of modulation in $^1$D, retention times in $^2$D will be subject to non-random shifts (intra-sample shift) because analytes eluting over several modulations will travel faster through the second column (Figure 3A).[20] Consequently, the PARAFAC model will be biased and more flexible alternatives such as flexible-coupling-PARAFAC2[13,34] or shift-invariant tri-linearity constrained MCR will likely give more accurate results.[28,35]

The situation becomes more complicated if the goal is to analyze a set of several GC×GC-TOFMS measurements jointly. In this situation, additional random shifts in $^1$D and $^2$D occur together with the intra-sample shifts in $^2$D (Figure 3B). Thus, shifts in both retention dimensions need to be accounted for.[20] To handle deviations from a multi-linear data structure, the PARAFAC2×N algorithm has been published as an extension to the flexible-coupling-PARAFAC2 model, which can effectively handle shifts in both retention dimensions.[13] However, the way algorithms from the "PARAFAC2-family" fundamentally address deviations from multi-linearity is rather complicated and come with assumptions regarding the nature of the shifts occurring in $^1$D and $^2$D that may have practical implications (see Supporting Information 1).[36,37] Therefore, a novel shift-invariant multilinearity constraint is proposed to account for shifts in both retention dimensions while providing unique solutions.

Conceptually, the shift-invariant multi-linearity constraint utilizes the shift-invariant property of the Fourier Transform modulus (amplitude spectra) to "de-shift" the estimates of elution profiles within the MCR-ALS routine to transform a non-multi-linear problem into a multi-linear problem, for which a unique solution exists.[28] The transition from non-multi-linear to a multi-linear problem is shown in Figure 2B-D, in which overlayed elution profiles (total ion chromatograms, TIC) of one analyte occurring in different samples are shown. By calculating the FFT of the elution profiles shown in Figure 2B, the shift along $^1$D can be effectively removed by mapping the elution profiles to the same phase spectrum (compare Figure 2C). Analogously, the shift along $^2$D can be removed by calculating the FFT along $^2$D and mapping the elution profiles to the same phase spectrum. The data obtained by de-shifting the elution profiles, shown in Figure 2D, could be modeled by one multi-linear factor, because the mass spectra for the same analyte are approximately constant across samples. Hence, the blue and orange factors only show a scale difference, which is due to their concentration difference.

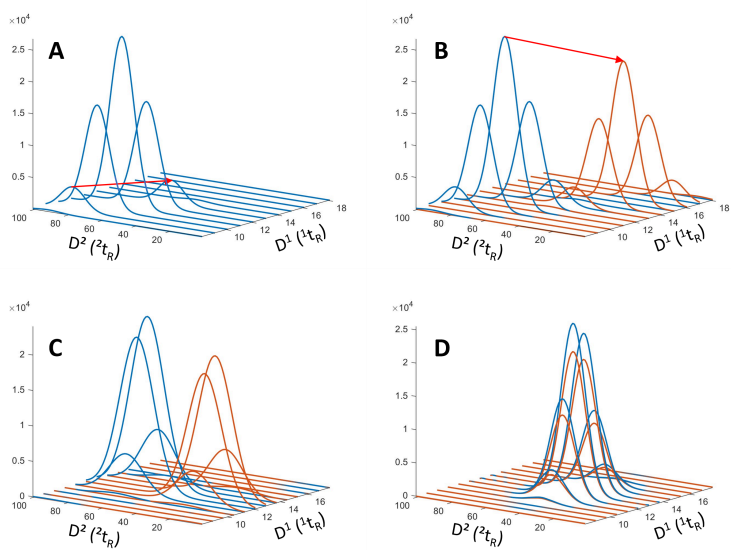

*Figure 2: Visualized are TIC profiles of one analyte present in two different samples (blue and orange peaks). **A:** Retention in $D^2$ gets faster with increasing modulation in $D^1$ because of higher temperatures on the second column (intra sample shift). **B:** Retention times between samples (plotted together as overlay) can vary because chromatographic conditions cannot be kept perfectly constant (inter sample shift), **C:** Synchronization of different peaks along $D^1$ after performing FFT and mapping the scans in $D^2$ to a common phase spectrum. **D:** full synchronization along $D^1$ and $D^2$ after performing FFT on the profiles shown in **C** along the modulations in $D^1$ and mapping the frequency spectra to a common phase spectrum.*

### 2.3 Shift-invariant multi-linearity

The shift-invariant multi-linearity constraint is implemented in the MCR-ALS routine as schematically shown in Figure 3. The example shows a simple case of a two-component system consisting of an analyte being present in multiple GC×GC-TOFMS measurements and a baseline signal. The steps for applying shift-invariant multi-linearity follow.

Step 1, the GC×GC-TOFMS measurements are rearranged from the fourth order tensor $\underline{X}^{(4)}$ ($I \times K \times L \times J$) to the augmented matrix $X$ ($IKL \times J$). The dimensions are the same as introduced in Section 2.1, $I$ denoting modulations in $^1$D, $K$ denoting scans in $^2$D, $L$ is the number samples and $J$ is the mass spectral dimension.

Step 2, the first estimates of the concatenated elution profiles and mass spectra $C_1$ and $S_1$ are obtained by regressing $S_0$ onto $X$ (to obtain $C_1$) and $C_0$ onto $X$ (to obtain $S_1$). Both regressions can be described as non-negative classical least square (CLS)[38] steps. To the estimates of $C_1$ are further constraints applied before a new iteration cycle is started (as described in the following Steps 3-8). In this notation, $S_0$ and $C_0$ are the initialized factor matrices and $C_i$ and $S_i$ with $i \in \{1, ..., I\}$ are estimates after the $i^{th}$ iteration. Positive random values are a straightforward choice for starting values, however, more sophisticated initialization schemes exist.[39,40]

Steps 3 through 8 apply the shift-invariant multi-linearity constraint sequentially to the concatenated elution profiles $c_{1,r}$ stored in $C_1$, with $r \in \{1, ..., R\}$ denoting the number of components. Step 3 extracts the $r^{th}$ elution profile, $c_{1,r}$.

In Step 4, the selected vector of elution profile estimates $c_{1,r}$ is smoothed using a wavelet transform denoiser, which is explained in more detail in the Supporting Information 2. Although, the denoising step in 4a is not strictly required for the multi-linearity, results discussed below show that it ensures very accurate estimates of mass spectra and elution profiles even at very low signal-to-noise ratios (SNR). Applying smoothing to the estimated elution profiles inside the ALS routine was found to be advantageous compared to smoothing of the raw data (compare Supporting Information 2). The combination of shift-invariant multi-linearity and wavelet based denoising will be distinguished from shift-invariant multi-linearity (SIML) by the abbreviation SIML-DN.

Step 4b reshapes $c_{1,r}$ into a matrix $C_{1,r}$ of dimension ($I \times KL$) and Step 4c synchronizes along $^1$D, using the FFT. After the synchronization, $C_{1,r}$ contains the amplitude spectra of the $^1$D elution profiles in its rows. The synchronization is achieved because the amplitude spectra (real part of the FT) of the elution profiles are shift-invariant, and the separately stored phase spectra contain the shift information (imaginary part of the FT). A more detailed explanation of this procedure can be found in Supporting Information 2 or in the literature.[41,42] Step 4d enforces shift-invariant multi-linearity on the elution profiles along $^1$D by reconstructing $C_{1,r}$ with a one-component SVD model.[28] The left-hand singular values contain the interim estimates of the $^1$D elution profiles in the frequency domain, while the right-hand singular values contain the concatenated $^2$D elution profiles which still need to be synchronized and enforced to follow the multi-linearity constraint. To achieve synchronization for the $^2$D elution profiles, Step 5a reshapes the right-hand singular values of the SVD from Step 4d to form a matrix of size $(K \times L)$. The synchronization procedure as described above in Steps 4c and 4d is repeated in steps 5b and 5c. The results of step 5c are the interim estimates of the $^2$D elution profiles in the frequency domain (left-hand singular values) and the relative concentrations of the analyte in the different samples (right hand singular values). For the reconstruction of the original, time domain elution profiles, the amplitude spectra of the $^1$D and $^2$D elution profiles are reconstructed by their 1-SVD-estimate in Step 6a and 7a. Afterwards, the inverse Fast Fourier Transform is applied in Step 6b and 7b to the multi-linearity constrained amplitude spectra and the separately stored phase spectra to obtain the shift-invariant-multi-linearity constrained time domain elution profiles. By that, the original peak positions are restored, and multi-linearity constrained elution profiles can be returned to finish the iteration and move on to the next cycle. As for conventional ALS routines, this procedure is repeated until a stop criterion is fulfilled. This could be that the changes in the loss function $L(C,S)$ are close to the numerical accuracy or that the defined number of maximum iterations is reached.

In summary, the trick behind shift-invariant multi-linearity is that multi-linearity is enforced on the shift-invariant amplitude spectra of the elution profiles rather than on the elution profiles themselves. In the conventional multi-linearity constraint, the $^1$D and $^2$D elution profiles of an analyte must be constant in position and shape across a set of samples up to their magnitude (resembling the concentration). Conversely, in shift-invariant multi-linearity it is the amplitude spectra of the respective elution profiles that must be constant across the samples, up to their magnitude. By reconstructing the matrices holding the amplitude spectra with their respective one-component SVD decomposition, multi-linearity is enforced. Because of the implementation of shift-invariant multi-linearity into the MCR-ALS routine, further constraints, and pre-processing steps (like the wavelet-based denoising) can easily be integrated.

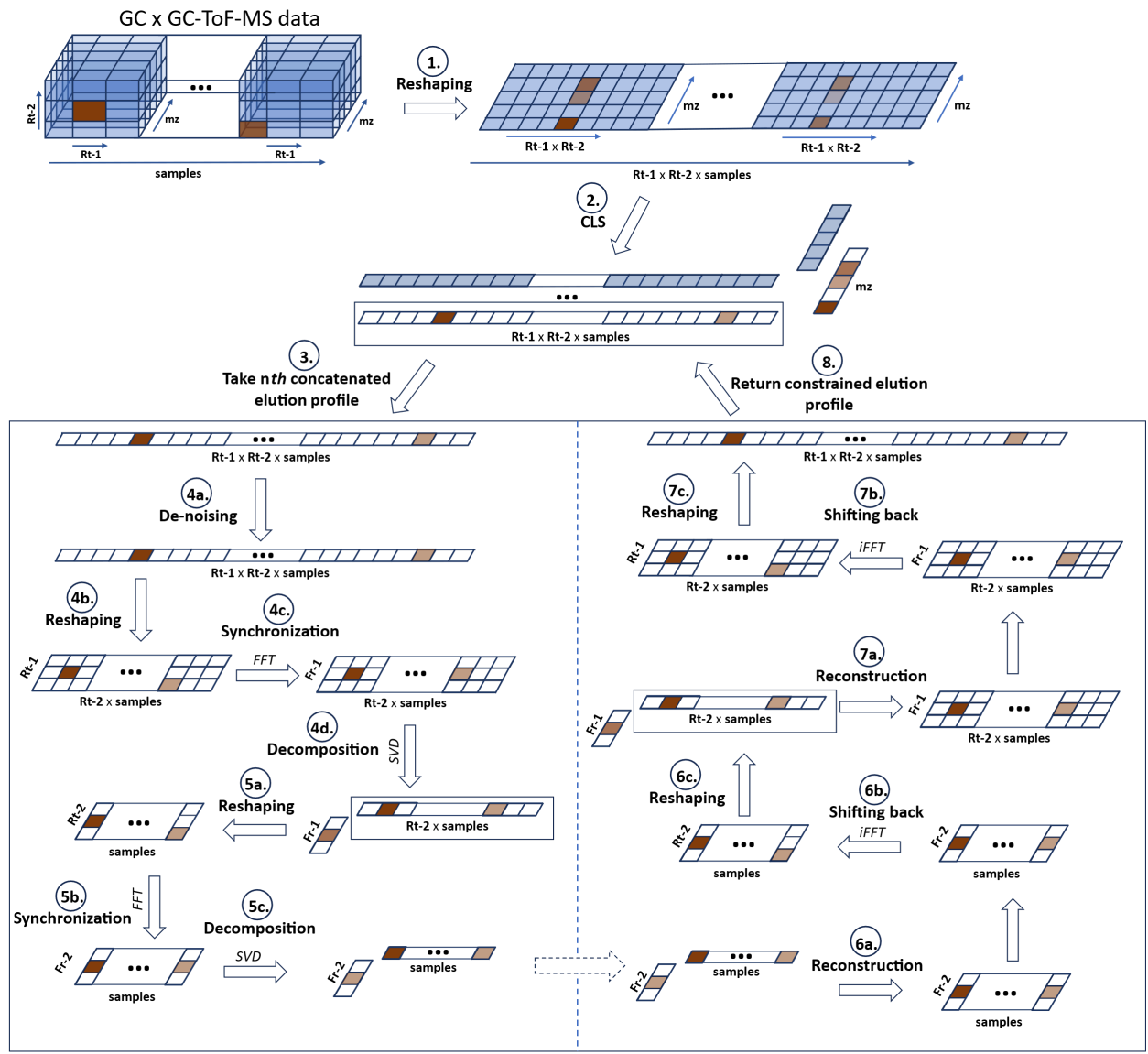

*Figure 3: Visualization of the shift-invariant multi-linearity algorithm with denoising. The estimates of the concatenated elution profiles are constraint after regressing the reshaped (step 1) GCxGC-TOFMS raw data on the estimated mass spectra using non-negative constraint CLS (step 2). Steps 3 to 8 show how the constraint is applied to the concatenated elution profiles.*

## 3. Material and methods

### 3.1 Software and algorithms

The performance of shift-invariant multi-linearity with and without denoising was benchmarked against the performance of non-negativity constrained MCR-ALS, and PARAFAC2×N. The same criterion for algorithmic convergence was applied, based on the relative change in the loss function value. The maximum number of iterations was set to 1.000 and it was assured that only converged models were used in the benchmark. All algorithms were initialized with positive random values, except PARAFAC2xN was initialized with a "best out of 10" (positive) random starts method. This was necessary to avoid an excessive number of local minima solutions.[39] Each algorithm was fitted multiple times starting from different random values. In total 50 converged models per algorithm and data set were compared to get an estimate of the precision and stability of the different algorithms.

MATLAB version R2022b (The MathWorks, Natick, MA USA) has been used for implementing the shift-invariant multi-linearity algorithm and the MCR-ALS routines. The PARAFAC2×N algorithm was taken from the GitHub repository: https://github.com/mdarmstr/parafac2x2, 24.03.2024. Area of feasible solution calculation were executed using the FACPAC software version 2.0 (http://www.math.uni-rostock.de/facpack/Downloads.html#Current_release, 24.03.2024)[43].

### 3.2 Simulated data

Data has been simulated to mimic a GC×GC-TOFMS data set with well-defined SNR using the function developed by Sorochan Armstrong et al[13]. The source code of the function is available under the URL https://github.com/mdarmstr/parafac2x2, 05.03.2024. Specifically, seven data sets emulating repeated measurements of two analytes eluting in retention-window ($^1t_R \times {}^2t_R$) were simulated under different SNR conditions. The number of samples in each data set was set to 10 and the number of modulations in $^1D$ was set to 20 and the number of scans in $^2D$ was set to 200. The mass axis was clipped to a range of 761 $m/z$ values. To study the performance of the algorithm in different SNR regimes, the SNR was varied from 3 to 0.025. Figure 3 shows examples of simulated measurements the data sets with moderate SNR and with low SNR. While for the SNR of 0.1 peak shapes are visible in the contour plots of the 2D-TIC and the 2D-BPCs, at the low SNR hardly any peak-like structure is recognizable.

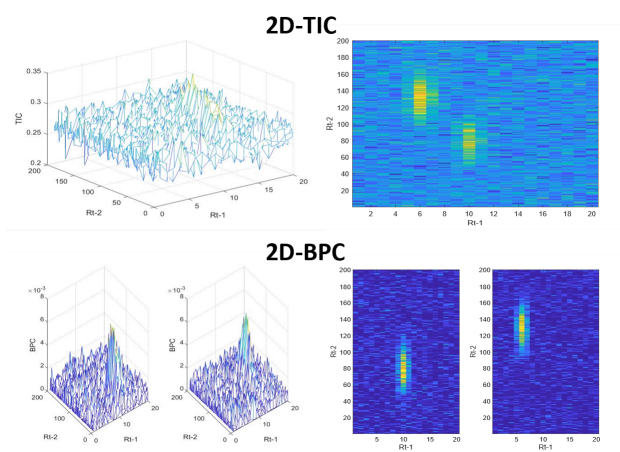
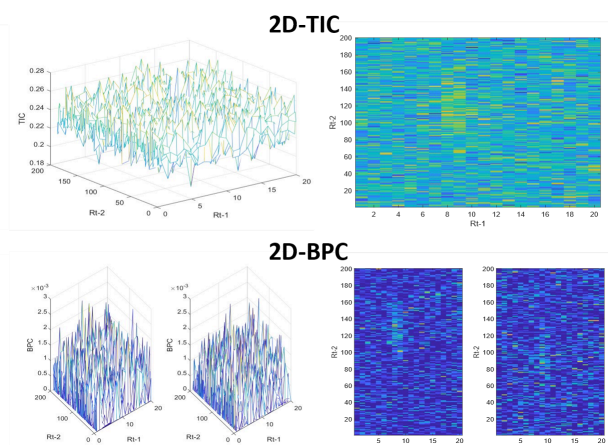

*Figure 4: **A:** Example of one simulated measurement at SNR of 0.1. **B:** Example of one simulated measurement at SNR of 0.025. The total ion chromatogram and the base peak chromatogram are shown to illustrate the noisiness of the simulated data.*

### 3.3 Experimental data

Previously published GC×GC-TOFMS data from a calibration experiment was used to compare the algorithm performance on real data. The two analytes present in the modeled ($^1t_R \times {^2t_R}$)-frame are the derivatized forms of salicylic acid and adipic acid (both molecules are derivatized with two TMS groups each, to cover the acid and hydroxy functionalities). For details on the derivatization procedure, we refer the reader to the original article.[13]

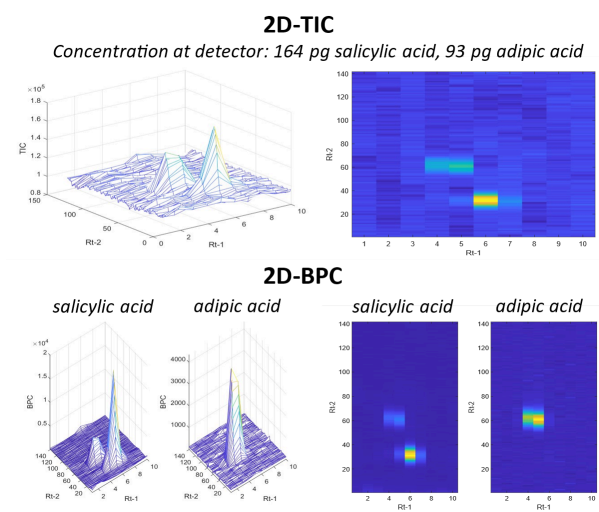
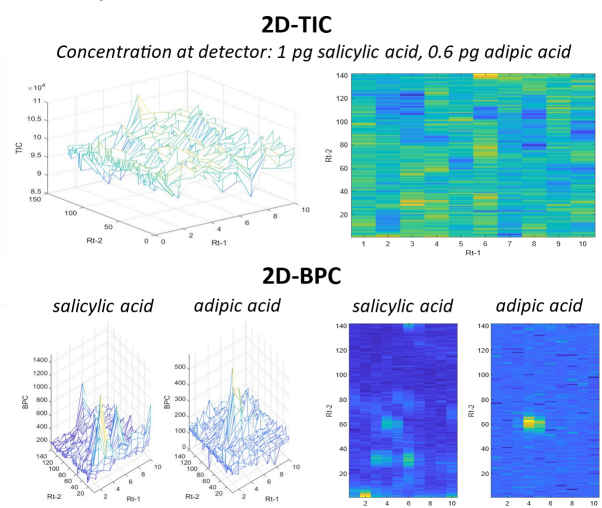

*Figure 5: **A:** Example of one of the triplicates at the highest calibration concentration considered (calibration point 6 in Table 1). The total ion chromatogram and the base peak chromatogram for salicylic and adipic acid are shown. The base peak chromatogram of salicylic acid is a fragment shared by adipic acid. **B:** Example of one of the triplicates at a lower calibration concentration (calibration point 12 in Table 1). The total ion chromatogram and the base peak chromatograms show the*

*noisiness of the data at the low concentration. The noise is different from the simulated case as it has a lower frequency and is more structured.*

The original data set consists of 14 calibration points covering a range of injected analyte amount of 0.1 pg – 16,393 pg for salicylic acid and 0.1 – 9,311 pg for adipic acid on column. Each calibration point has been measured in triplicates. Since we were mostly interested in investigating algorithm performance in the low SNR domain, we discarded the first five calibration points from the data set to begin with and then continuously removed further calibration points to study the breakdown point for each algorithm. The highest calibration point in the first data set is 163.9 pg salicylic acid and 93.1 pg adipic acid and the highest calibration point in the last data set is 1 pg salicylic acid and 0.6 pg adipic acid.

*Table 1: List of the calibration standards published by Armstrong et al. with their respective concentrations on column in pg. Standards 1-5 have been removed from the set for this study and standard 6 has been used to determine the precision in the extrapolation experiment (see description in 3.4).*

| Cal. standard | Salicylic acid [pg] | Adipic acid [pg] | Replicates |
|---|---|---|---|
| 1* | 16392.6 | 9311.1 | 3 |
| 2* | 8196.3 | 4655.6 | 3 |
| 3* | 2732.1 | 1551.9 | 3 |
| 4* | 1092.8 | 620.7 | 3 |
| 5* | 327.9 | 186.2 | 3 |
| 6** | 163.9 | 93.1 | 3 |
| 7 | 82.0 | 46.6 | 3 |
| 8 | 27.3 | 15.5 | 3 |
| 9 | 10.9 | 6.2 | 3 |
| 10 | 5.5 | 3.1 | 3 |
| 11 | 2.7 | 1.6 | 3 |
| 12 | 1.0 | 0.6 | 3 |
| 13 | 0.5 | 0.3 | 3 |
| 14 | 0.1 | 0.1 | 3 |

### 3.4 Metrics for evaluation

The performance of the algorithms was assessed using different quantitative metrics. The fit measured as variance explained ($VarExpl$) was used to evaluate how good the different methods can explain the data (Equation 4-6). In Equation 5-6, $x_{n,m}$ is the entry in the $n^{th}$ row and $m^{th}$ column of the original data matrix $X$ ($IKL \times J$) and $\hat{x}_{n,m}$ is the entry in the $n^{th}$ row and $m^{th}$ column of the matrix $\hat{X}$ ($IKL \times J$), reconstructed from $C$ and $S^T$.

Equation 4:
$$VarExpl = \left(1 - \frac{SSE}{SST}\right) * 100$$

where

Equation 5:
$$SSE = \sum_{n}^{IKL} \sum_{m}^{J} (x_{n,m} - \hat{x}_{n,m})^2$$

and

Equation 6:
$$SST = \sum_{n}^{IKL} \sum_{m}^{J} (x_{n,m})^2$$

The cosine similarity (also known as Tucker congruence) was calculated according to Equation 7 to assess how well the estimated mass spectra correspond to the true mass spectra. In Equation 7, $s_{r,est}$ is the estimated mass spectrum and $s_{r,ref}$ is the reference mass spectrum.

Equation 7:
$$cosine\ similarity = \frac{s_{r,est}^T s_{r,ref}}{\|s_{r,est}\|_F^2 \|s_{r,ref}\|_F^2}$$

The quality of the estimated concentrations was assessed in two different ways for the simulated data and for the real data. For the simulated data, calibration curves were fitted between the estimated concentrations and the true concentrations. The $R^2$ value and the bias (offset) of the calibration curve were evaluated as they are commonly used figures of merit in quantitative chromatographic analysis.

For the real calibration data, an extrapolation experiment was performed, in which the concentration of the highest calibration point (163.9 pg salicylic acid and 93.1 pg adipic acid) was successively predicted with models which were built on the $(9 - p)$ lowest calibration points. The pooled, relative standard deviation was used to assess the quantitative precision of the different methods across the fitted models after removing $p = 1, ..., 6$ calibration standards.

## 4. Results and Discussion

### 4.1 Simulated Data

The results of using the different models on the simulated data set indicate larger performance differences between MCR-ALS, SIML, and PARAFAC2×N, which are summarized in Figure 6 and Figure 7. While MCR-ALS and SIML achieve nearly the same fit values on the simulated data sets, PARAFAC2×N fits the data on average significantly worse. Moreover, Figure 6A also shows that the fit values achieved with PARAFAC2×N are subject to larger variation across the 50 repeated fits, which indicates that some of the solutions converged to local minima.[39] The performance difference between PARAFAC2×N and the other models becomes more pronounced when comparing accuracies of the estimated mass spectra at different noise levels (Figure 6B). The cosine similarity between the estimated and the underlying true spectra declines rapidly for all algorithms when the SNR is lower than 0.5.

In direct comparison, at SNR of 0.1 SIML has cosine similarity (Figure 6B) and $R^2$ values (Figure 6C) close to one but for MCR the performance starts to degrade. The performance of SIML is slightly better than MCR at SNR of 0.05 and similar to MCR at SNR of 0.025. In contrast, SIML-DN performed well at all noise levels studied and provided high cosine similarity and $R^2$ values even at an SNR of 0.025.

With respect to the calibration curves it is however noticeable, that SIML with denoising has a larger bias than MCR-ALS and SIML without denoising at higher SNRs (0.5 - 3). This difference vanishes at lower SNR values (0.025 - 0.1), at which the bias of MCR-ALS, PARAFAC2×N and SIML without denoising becomes larger than the bias of SIML with denoising. The reason for this is that the models without denoising have increasing difficulties separating the baseline from the analyte signal, which can also be emphasized by comparing the ²D elution profiles and mass spectra shown in Figure 7 at SNR of 0.1 and 0.025. Especially, the comparison of the elution profiles and mass spectra at a SNR of 0.025 highlight the stability of the SIML-DN method, because it still provides reasonable estimates whereas the estimates of the other methods can hardly be qualified as chemical information.

Although SIML appears to be more stable than MCR-ALS based on the results of the simulation study, the differences are not huge, supporting MCR-ALS as a strong benchmark.

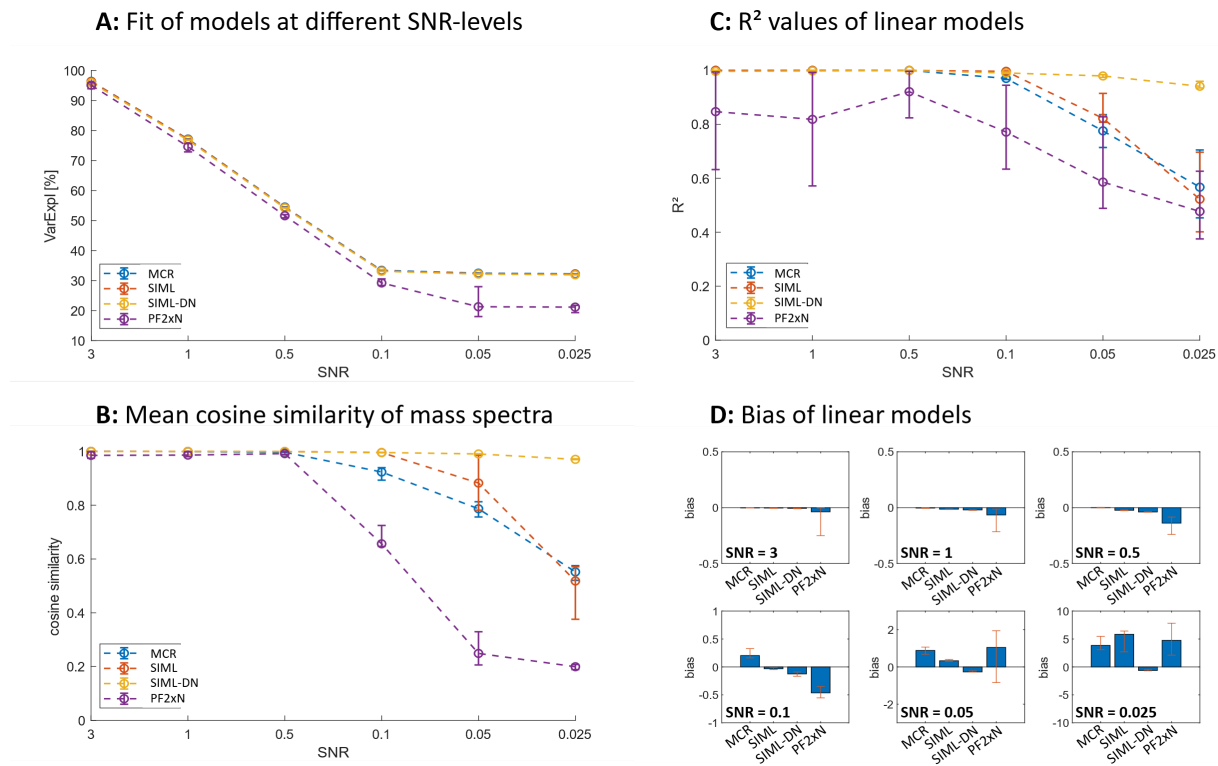

*Figure 6: Summary of the performance of MCR, SIML, SIML-DN and PF2×N on simulated GC×GC-TOFMS data at different SNRs. In all plot are the mean values and standard deviations over 50 repeated fits at each SNR visualized. **A:** variance explained indicating how well the different models describe the data. All models show the same trend that as the SNR descreases, the fit of the models gets worse. PF2xN shows significantly lower fit compared to all the other models. **B:** Mean cosine similarity of the estimated mass spectra and the true mass spectra. The models show distinct capability of modelling the true mass spectra at different SNR. The SIML-DN algorithm is most robust against high noise levels. **C:** $R^2$ values of linear models fitted on the peak areas and the known concentration values. The SIML-DN algorithm is most robust against high noise levels and PF2xN shows higher variability, probably due to the presence of local minima. **D:** Bias of linear models fitted on the peak areas and the known concentration values. The bias of MCR, SIML and PF2xN increases more with decreasing SNR than the bias of SIML-DN*

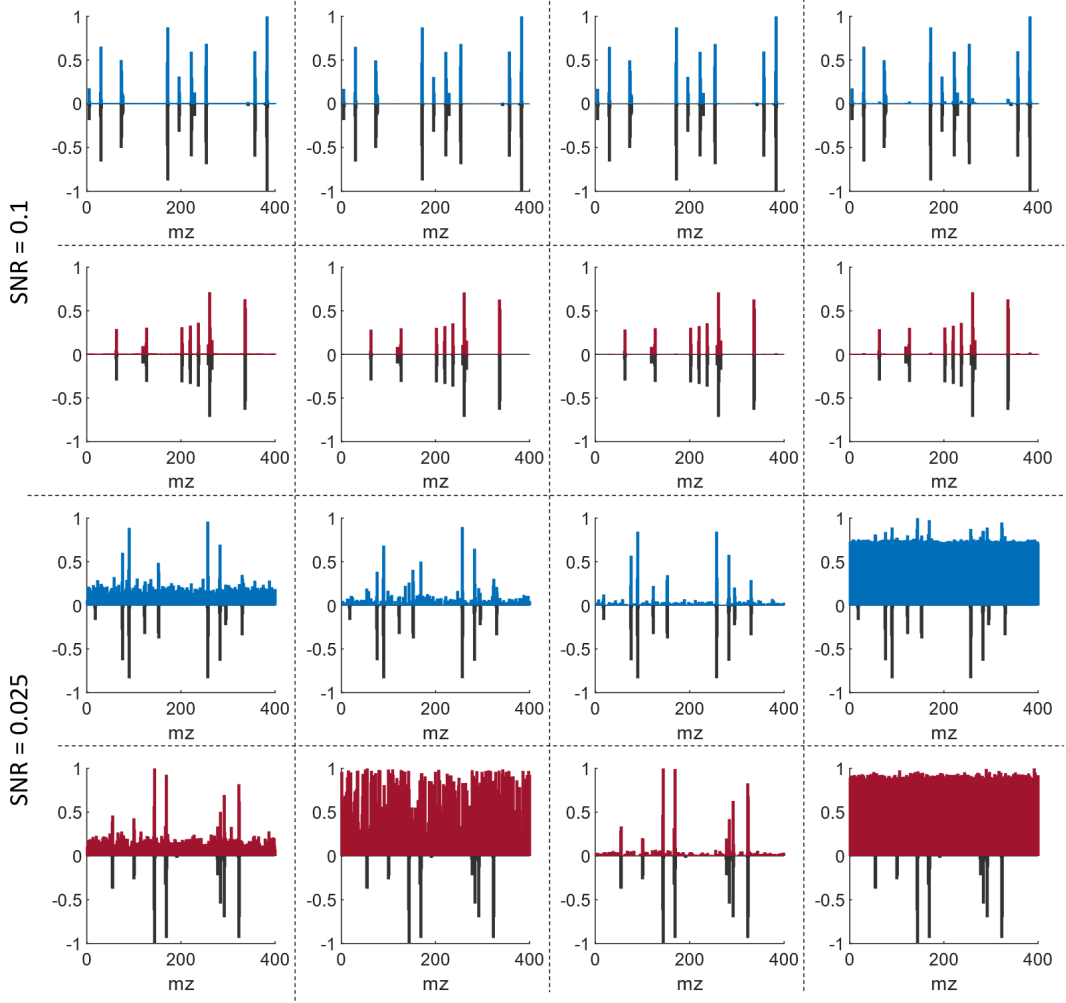
Internal

*Figure 7: Comparison of estimated elution profiles and mass spectra at high and at low SNR. The colored elution profiles and spectra resemble estimates, while dashed and solid black lines show the true reference profiles and spectra.*

### 4.2 Experimental Data

In comparison to the results for the simulated data, the situation changes quite dramatically when looking at the calibration data of salicylic and adipic acid. Although MCR-ALS fits the data better than SIML and SIML-DN (Figure 8A), the cosine similarities between the estimated and the true spectra are on average significantly worse than the estimates of SIML and SIML-DN (Figure 8B). Moreover, the rotational ambiguity of the MCR model translates to a larger variability in the estimated spectral profiles compared to SIML and SIML-DN. The same holds for the precision of the predictions of the calibration standard (163.9 pg salicylic acid / 93.1 pg adipic acid) shown in Figure 8C. The relative error is calculated from the predictions made with models built on the calibration data after removing 6, 7, … ,11 calibration points, as pointed out in section 3.4. In the extremes, a model was built on calibration data ranging from 0.1 to 1.1 pg and from 0.1 to 0.6 pg for salicylic acid and adipic acid, respectively, to predict concentrations of 163.9 pg salicylic acid and 93.1 pg adipic acid.

While the inter quartile range of the relative prediction error from the MCR models reaches from -0.5 to 0.1 (salicylic acid) and from -0.5 to 0.5 (adipic acid), the inter quartile range for the SIML and SIML-DN models reaches at most from -0.02 to 0.02 considering salicylic acid and adipic acid. The few high prediction errors from the SIML models for adipic acid can be related to the results from models built on calibration data after removing 11 standards. At this point the SIML models reach their break down point (compare Figure 8B), while SIML-DN still provides reliable mass spectra and concentration estimates.

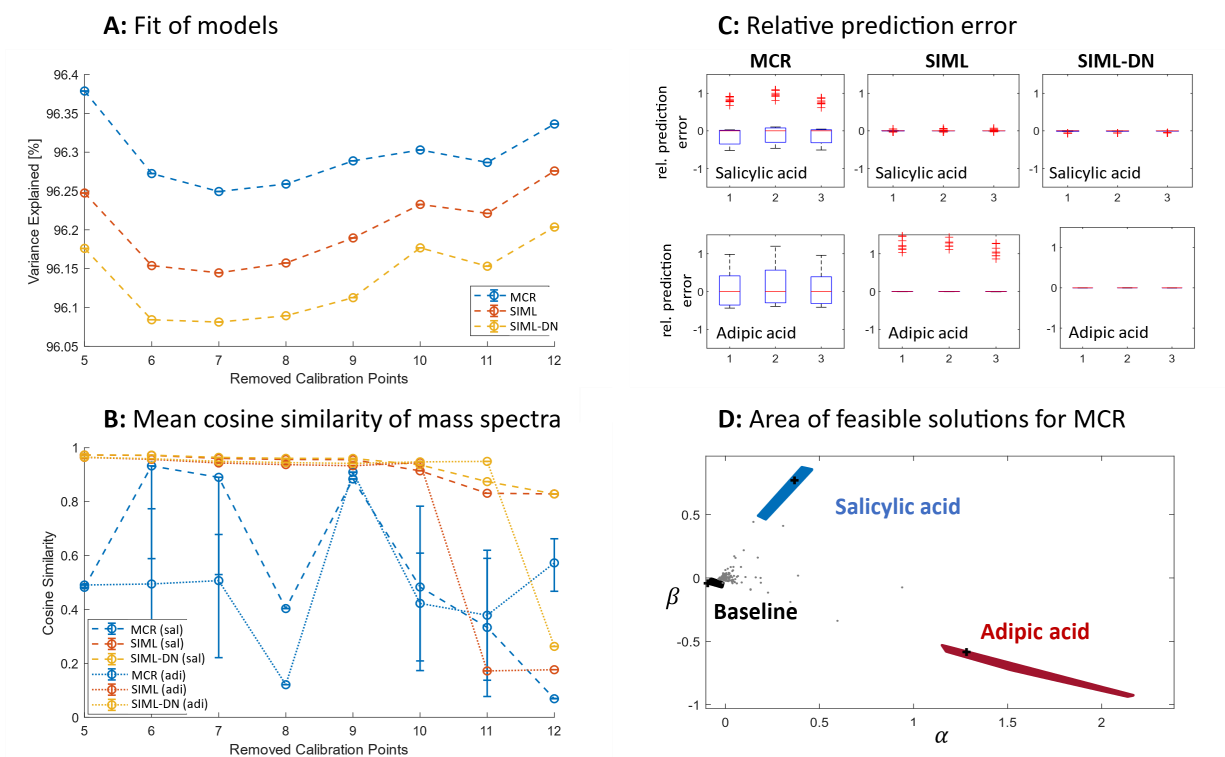

*Figure 8: Summary of the performance of MCR, SIML and SIML-DN on a real GC×GC-TOFMS calibration data set. In all plot are the mean values and standard deviations over 50 repeated fits on increasingly smaller subsets of the calibation data visualized. **A:** variance explained indicating how well the different models describe the data. All models show the same trend, however MCR tends to have silghtly better fit, followed by SIML and SIML-DN has the lowest fit. This trend follows the intuition that the least constrained model should have the highest fit. **B:** Mean cosine similarity of the estimated mass spectra and the true mass spectra. The estimates obtained from MCR show large variance which can be accounted to rotational ambuigity. The estimates of SIML and SIML-DN show high accuracy up to the removal of 11 and 12 calibration points, respectively. **C:** Relative prediction error that is made when the highest calibration standard is predicted with models trained on subsets of the calibration data set after removing up to 11 calibration points. The relative prediction error made with the MCR models is in the order of magnitude of ± 50 % while the relative prediction error for SIML and SIML-DN is < ± 2 % up to the point when 11 calibration points are removed. At this point the prediction error of the SIML model increases drastically for adipic acid. **D:** Area of feasible solution for MCR on the whole concatenated calibration data set. The area of feasible solution shows that MCR suffers substantially from rotational ambiguity, even on the data set including all calibration points.*

This big difference in the results is because MCR is suffering from rotational ambiguity (see Figure 8D) whereas SIML and SIML-DN provide unique solutions. Compared to the simulated data, the calibration data set is a harder challenge because the mass spectra as well as the concentration profiles are correlated. The correlation in the mass spectra can be traced back to shared fragments that stem from the TMS derivatization (specifically fragments 73,74, and 147).[44] Derivatization with TMS or other reagents is a common practice in gas chromatography and will may cause similar problems for MCR because the mass spectra will contain shared fragments. The correlation in the concentration profiles is straightforward to explain because the calibration data represents a dilution series. In Figure 9 are the $^2$D elution profiles and mass spectra shown from models built on calibration data after removing 6 and 11 calibration standards, respectively. In all cases the models that achieved the highest cosine similarity with their spectral estimates were selected for visualization in Figure 9. Among these models, the difference between the MCR and the SIML estimates is not very pronounced but still visible in the offset of the MCR elution profiles at the higher concentration level. The visualization of the estimates from the smallest calibration set shows once more that the denoising implemented in SIML-DN pays off because it allows for the extraction of useful chemical information under really challenging conditions.

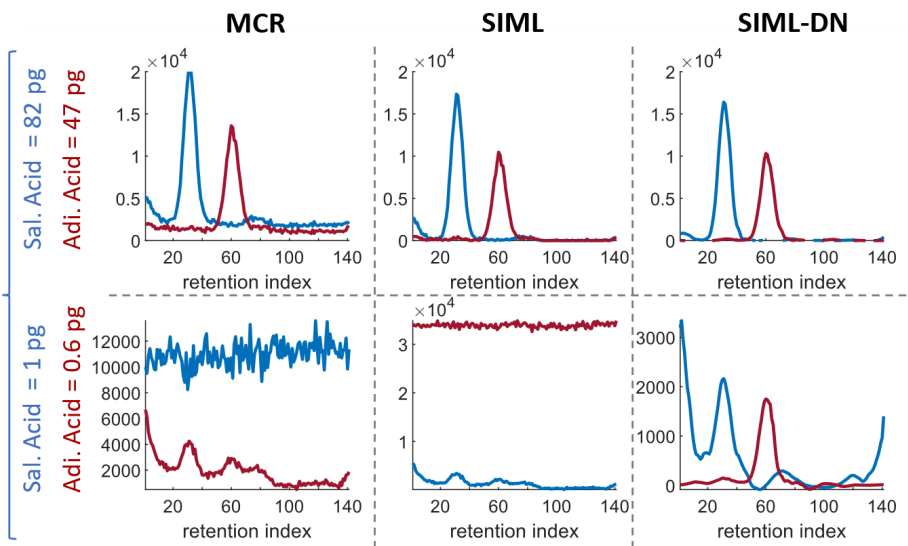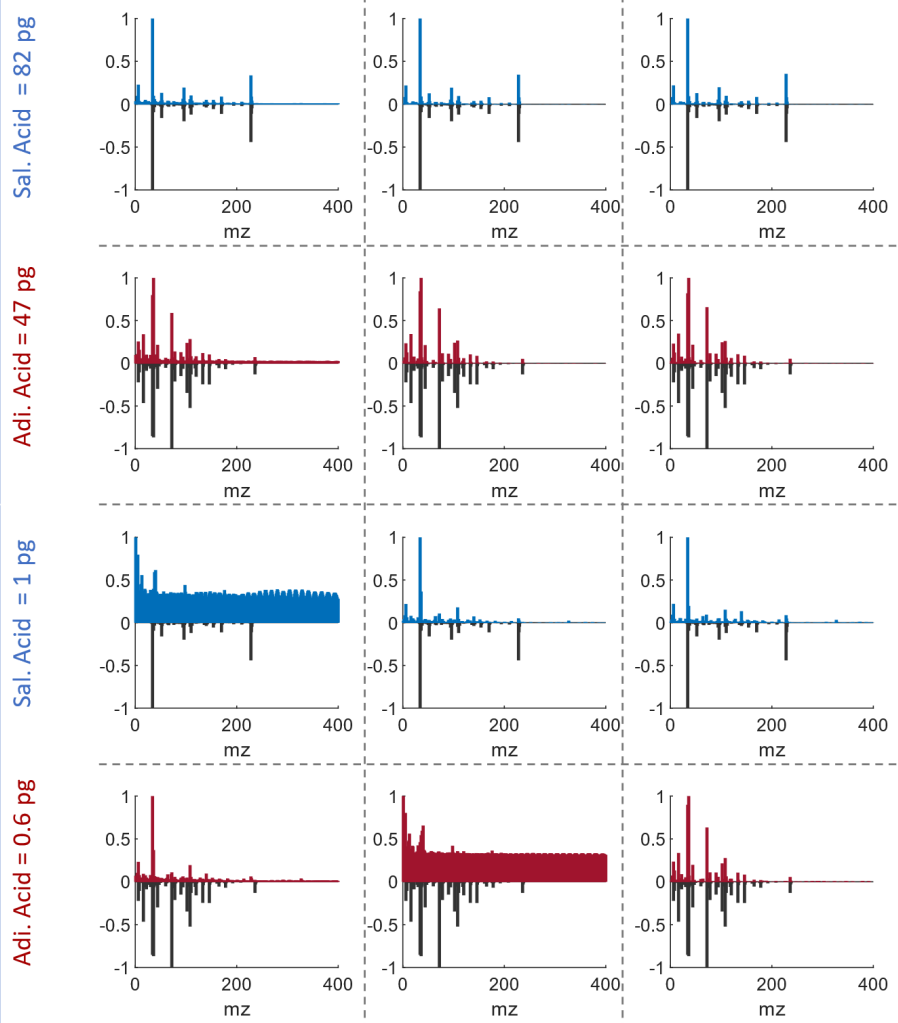

*Figure 9: Comparison of estimated elution profiles and mass spectra at high and at low SNR. The colored elution profiles and spectra resemble estimates, while dashed and solid black lines show the true reference profiles and spectra.*

## 5. Conclusions

In conclusion, the proposed shift-invariant multi-linearity (SIML) algorithm demonstrates a significant advancement in the analysis of comprehensive GC×GC-TOFMS data. By incorporating wavelet-based smoothing and Fourier-Transform based de-shifting into the multivariate curve resolution-alternating least squares (MCR-ALS) routine, the SIML algorithm successfully addresses the challenges of retention time shifts and high noise levels. Benchmarking against standard MCR-ALS and PARAFAC2×N methods reveals that SIML provides unique solutions and exhibits unmatched robustness against high noise levels, achieving impressive performance in both simulated and real data scenarios. The robustness is especially evident in lower signal-to-noise ratio (SNR) regimes, where SIML-DN maintains high accuracy in estimating mass spectra and concentrations. This enhances the reliability of compound identification and quantitative analyses in complex matrices, which is crucial for advancing the applications of GC×GC-TOFMS in various fields such as environmental science, food chemistry, and biological research.

## 6. Acknowledgements

The authors would like to express their gratitude to BASF SE for funding the PhD project that formed the basis of this research.

## 7. References

1.  Muscalu, A. M. & Górecki, T. Comprehensive two-dimensional gas chromatography in environmental analysis. *TrAC - Trends in Analytical Chemistry* vol. 106 225–245 Preprint at https://doi.org/10.1016/j.trac.2018.07.001 (2018).
2.  Yu, M., Yang, P., Song, H. & Guan, X. Research progress in comprehensive two-dimensional gas chromatography-mass spectrometry and its combination with olfactometry systems in the flavor analysis field. *Journal of Food Composition and Analysis* vol. 114 Preprint at https://doi.org/10.1016/j.jfca.2022.104790 (2022).
3.  Thong, A., Basri, N. & Chew, W. Comparison of untargeted gas chromatography-mass spectrometry analysis algorithms with implications to the interpretation and putative identification of volatile aroma compositions. *J Chromatogr A* **1713**, (2024).
4.  Zanella, D. *et al.* The contribution of high-resolution GC separations in plastic recycling research. *Analytical and Bioanalytical Chemistry* vol. 415 2343–2355 Preprint at https://doi.org/10.1007/s00216-023-04519-8 (2023).
5.  Stilo, F., Bicchi, C., Reichenbach, S. E. & Cordero, C. Comprehensive two-dimensional gas chromatography as a boosting technology in food-omic investigations. *Journal of Separation Science* vol. 44 1592–1611 Preprint at https://doi.org/10.1002/jssc.202100017 (2021).
6.  Reichenbach, S. E., Tao, Q., Cordero, C. & Bicchi, C. A data-challenge case study of analyte detection and identification with comprehensive two-dimensional gas chromatography with mass spectrometry (GC×GC-MS). *Separations* **6**, (2019).
7.  Prebihalo, S. E. *et al.* Multidimensional Gas Chromatography: Advances in Instrumentation, Chemometrics, and Applications. *Analytical Chemistry* vol. 90 505–532 Preprint at https://doi.org/10.1021/acs.analchem.7b04226 (2018).
8.  Prebihalo, S. E., Reaser, B. C. & Gough, D. V. Multidimensional Gas Chromatography: Benefits and Considerations for Current and Prospective Users. *LCGC North America* 508–513 (2022) doi:10.56530/lcgc.na.zi3478f2.
9.  Weggler, B. A. *et al.* A unique data analysis framework and open source benchmark data set for the analysis of comprehensive two-dimensional gas chromatography software. *J Chromatogr A* **1635**, (2021).
10. Stefanuto, P. H., Smolinska, A. & Focant, J. F. Advanced chemometric and data handling tools for GC×GC-TOF-MS: Application of chemometrics and related advanced data handling in chemical separations. *TrAC - Trends in Analytical Chemistry* vol. 139 Preprint at https://doi.org/10.1016/j.trac.2021.116251 (2021).
11. Furbo, S., Hansen, A. B., Skov, T. & Christensen, J. H. Pixel-Based Analysis of Comprehensive Two-Dimensional Gas Chromatograms (Color Plots) of Petroleum: A Tutorial. *Anal Chem* **86**, 7160–7170 (2014).
12. Sudol, P. E., Ochoa, G. S. & Synovec, R. E. Investigation of the limit of discovery using tile-based Fisher ratio analysis with comprehensive two-dimensional gas chromatography time-of-flight mass spectrometry. *J Chromatogr A* **1644**, 462092 (2021).
13. Sorochan Armstrong, M. D., Hinrich, J. L., de la Mata, A. P. & Harynuk, J. J. PARAFAC2×N: Coupled decomposition of multi-modal data with drift in N modes. *Anal Chim Acta* **1249**, (2023).
14. Zhang, Z., Ma, P. & Lu, H. Two-Way Data Analysis: Multivariate Curve Resolution: Noniterative Resolution Methods. in *Comprehensive Chemometrics* 137–152 (Elsevier, 2020). doi:10.1016/B978-0-12-409547-2.14875-9.
15. Lawton, W. H. & Sylvestre, E. A. Self Modeling Curve Resolution. *Technometrics* **13**, 617 (1971).
16. de Juan, A., Rutan, S. C. & Tauler, R. Two-Way Data Analysis: Multivariate Curve Resolution – Iterative Resolution Methods. in *Comprehensive Chemometrics* 325–344 (Elsevier, 2009). doi:10.1016/B978-044452701-1.00050-8.

17. Lin, C.-J. Projected Gradient Methods for Nonnegative Matrix Factorization. *Neural Comput* **19**, 2756–2779 (2007).
18. Paatero, P. & Tapper, U. Positive matrix factorization: A non-negative factor model with optimal utilization of error estimates of data values. *Environmetrics* **5**, 111–126 (1994).
19. Paatero, P. The Multilinear Engine: A Table-Driven, Least Squares Program for Solving Multilinear Problems, including the n-Way Parallel Factor Analysis Model. *Journal of Computational and Graphical Statistics* **8**, 854 (1999).
20. Parastar, H., Radović, J. R., Bayona, J. M. & Tauler, R. Solving chromatographic challenges in comprehensive two-dimensional gas chromatography-time-of-flight mass spectrometry using multivariate curve resolution-alternating least squares ABC Highlights: Authored by Rising Stars and Top Experts. *Anal Bioanal Chem* **405**, 6235–6249 (2013).
21. Hoggard, J. C. & Synovec, R. E. Parallel factor analysis (PARAFAC) of target analytes in GC × GC-TOFMS data: Automated selection of a model with an appropriate number of factors. *Anal Chem* **79**, 1611–1619 (2007).
22. Sawall, M., Schröder, H., Meinhardt, D. & Neymeyr, K. On the Ambiguity Underlying Multivariate Curve Resolution Methods. in *Comprehensive Chemometrics* 199–231 (Elsevier, 2020). doi:10.1016/B978-0-12-409547-2.14582-2.
23. Tauler, R., Smilde, A. & Kowalski, B. Selectivity, local rank, three-way data analysis and ambiguity in multivariate curve resolution. *J Chemom* **9**, 31–58 (1995).
24. Olivieri, A. C. A down-to-earth analyst view of rotational ambiguity in second-order calibration with multivariate curve resolution − a tutorial. *Anal Chim Acta* **1156**, 338206 (2021).
25. Bro, R. PARAFAC. Tutorial and applications. *Chemometrics and Intelligent Laboratory Systems* **38**, 149–171 (1997).
26. Bro, R., Andersson, C. A. & Kiers, H. A. L. PARAFAC2—Part II. Modeling chromatographic data with retention time shifts. *J Chemom* **13**, 295–309 (1999).
27. Tauler, R. Multivariate curve resolution of multiway data using the multilinearity constraint. *J Chemom* **35**, (2021).
28. Schneide, P. A., Bro, R. & Gallagher, N. B. Shift-invariant tri-linearity—A new model for resolving untargeted gas chromatography coupled mass spectrometry data. *J Chemom* **37**, (2023).
29. Van Benthem, M. H. & Keenan, M. R. Fast algorithm for the solution of large-scale non-negativity-constrained least squares problems. *J Chemom* **18**, 441–450 (2004).
30. Jaumot, J. & Tauler, R. MCR-BANDS: A user friendly MATLAB program for the evaluation of rotation ambiguities in Multivariate Curve Resolution. *Chemometrics and Intelligent Laboratory Systems* **103**, 96–107 (2010).
31. Olivieri, A. C., Neymeyr, K., Sawall, M. & Tauler, R. How noise affects the band boundaries in multivariate curve resolution. *Chemometrics and Intelligent Laboratory Systems* **220**, 104472 (2022).
32. Harshman, R. A. & Lundy, M. E. The PARAFAC model for three-way factor analysis and multidimensional scaling. in *Research methods for multimode data analysis* (eds. Law, H. G., Snyder Jr, C. W., Hattie, J. A. & McDonald, R. P.) 122–215 (Praeger, New York, 1984).
33. Carroll, J. D. & Chang, J.-J. Analysis of individual differences in multidimensional scaling via an n-way generalization of "Eckart-Young" decomposition. *Psychometrika* **35**, 283–319 (1970).
34. Cohen Jeremy E. and Bro, R. Nonnegative PARAFAC2: A Flexible Coupling Approach. in *Latent Variable Analysis and Signal Separation* (ed. Deville Yannick and Gannot, S. and M. R. and P. M. D. and W. D.) 89–98 (Springer International Publishing, Cham, 2018).
35. Zhang, X. & Tauler, R. Flexible Implementation of the Trilinearity Constraint in Multivariate Curve Resolution Alternating Least Squares (MCR-ALS) of Chromatographic and Other Type of Data. *Molecules* **27**, 2338 (2022).

36. Cohen Jeremy E. and Bro, R. Nonnegative PARAFAC2: A Flexible Coupling Approach. in *Latent Variable Analysis and Signal Separation* (ed. Deville Yannick and Gannot, S. and M. R. and P. M. D. and W. D.) 89–98 (Springer International Publishing, Cham, 2018).
37. Kiers, H. A. L., ten Berge, J. M. F. & Bro, R. PARAFAC2—Part I. A direct fitting algorithm for the PARAFAC2 model. *J Chemom* **13**, 275–294 (1999).
38. Gallagher, N. B. Classical least squares for detection and classification. in 231–246 (2019). doi:10.1016/B978-0-444-63977-6.00011-0.
39. Yu, H. & Bro, R. PARAFAC2 and local minima. *Chemometrics and Intelligent Laboratory Systems* **219**, 104446 (2021).
40. Nardecchia, A. & Duponchel, L. Randomised SIMPLISMA: Using a dictionary of initial estimates for spectral unmixing in the framework of chemical imaging. *Talanta* **217**, 121024 (2020).
41. Fourier and Wavelet Transforms. in *Data-Driven Science and Engineering* 47–83 (Cambridge University Press, 2019). doi:10.1017/9781108380690.003.
42. Cooley, J. W. & Tukey, J. W. An algorithm for the machine calculation of complex Fourier series. *Math Comput* **19**, 297–301 (1965).
43. Sawall, M. & Neymeyr, K. A fast polygon inflation algorithm to compute the area of feasible solutions for three-component systems. II: Theoretical foundation, inverse polygon inflation, and *FAC-PACK* implementation. *J Chemom* **28**, 633–644 (2014).
44. DeJongh, D. C. *et al.* Analysis of trimethylsilyl derivatives of carbohydrates by gas chromatography and mass spectrometry. *J Am Chem Soc* **91**, 1728–1740 (1969).

Internal